\documentclass[journal,12pt,onecolumn,draftclsnofoot,]{IEEEtran}
\IEEEoverridecommandlockouts
\usepackage{cite}
\usepackage{amsmath,amssymb,amsfonts}
\usepackage{algorithmic}
\usepackage{multirow}
\usepackage[pdftex]{graphicx}
\usepackage{textcomp}
\def\BibTeX{{\rm B\kern-.05em{\sc i\kern-.025em b}\kern-.08em
    T\kern-.1667em\lower.7ex\hbox{E}\kern-.125emX}}
\usepackage{color}
\usepackage{tikz,pgfplots}
\pgfplotsset{compat=newest}

\begin{document}

\title{Mapping of Local and Global Synapses on Spiking Neuromorphic Hardware}

\author{\IEEEauthorblockN{Anup Das${}^{\ddagger\dagger}$, Yuefeng Wu${}^\dagger$, Khanh Huynh${}^\dagger$, Francesco Dell'Anna${}^\mathsection$, Francky Catthoor${}^\mathsection$, Siebren Schaafsma${}^\dagger$}
\IEEEauthorblockA{\textit{${}^\dagger$Stichting IMEC Nederland}, 
\textit{${}^\mathsection$IMEC Belgium}, 
\textit{${}^\ddagger$Department of ECE, Drexel University, Philadelphia, USA}\\
Correspondance email: akdas@ieee.org}
}

\maketitle

\begin{abstract}
	Spiking Neural Networks (SNNs) are widely deployed to solve complex pattern recognition, function approximation and image classification tasks.  With the growing size and complexity of these networks, hardware implementation becomes challenging because scaling up the size of a single array (crossbar) of fully connected neurons is no longer feasible due to strict energy budget. Modern neromorphic hardware integrates small-sized crossbars with time-multiplexed interconnects. Partitioning SNNs becomes essential in order to map them on neuromorphic hardware with the major aim to reduce the global communication latency and energy overhead. To achieve this goal, we propose our instantiation of particle swarm optimization, which partitions SNNs into local synapses (mapped on crossbars) and global synapses (mapped on time-multiplexed interconnects), with the objective of reducing spike communication on the interconnect. This improves latency, power consumption as well as application performance by reducing inter-spike interval distortion and spike disorders. Our framework is implemented in Python, interfacing CARLsim, a GPU-accelerated application-level spiking neural network simulator with an extended version of Noxim, for simulating time-multiplexed interconnects. Experiments are conducted with realistic and synthetic SNN-based applications with different computation models, topologies and spike coding schemes. Using power numbers from in-house neuromorphic chips, we demonstrate significant reductions in energy consumption and spike latency over PACMAN, the widely-used partitioning technique for SNNs on SpiNNaker.

\end{abstract}


\section{Introduction}
\label{sec:introduction}
Spiking Neural Networks (SNNs) \cite{maass1997networks} are powerful and biologically realistic computation models, inspired by the dynamics of human brain. From implementation viewpoint, SNNs are collection of neurons that communicate by sending short pulses (spikes) across connections (synapses) to other neurons. SNNs are trained to perform variety of tasks, where the training process involves adjusting connection strengths between neurons. SNNs are increasingly being deployed to solve complex tasks such as pattern recognition, function approximation and image classification. Another reason for widespread success of SNNs are their efficient VLSI implementations, such as TrueNorth~\cite{akopyan2015truenorth}, CxQuad~\cite{indiveri2015neuromorphic} and SpiNNaker~\cite{khan2008spinnaker} among others. Our work is based on CxQuad, which is an analog neuromorphic hardware with 1024 neurons clustered into four crossbars of 256 neurons each. The framework proposed in this work can be extended with reasonable effort reusing the same concepts to other neuromorphic architectures.

With increasing deployment of SNNs in complex scenarios, the field is rapidly maturing in terms of applications, algorithms, computation models and hardware. Although significant research activities in these domains are being conducted separately, the task of a system designer is to build synergy between these domains i.e., to map realistic SNN-based applications on neuromorphic hardware. Although some efforts are made in this direction, the published approaches still remain mostly ad-hoc \cite{ji2016neutrams}, specific to a given hardware \cite{chen2017eyeriss} and are often limited to trivial applications \cite{serrano2009caviar}. The closest technique to our approach is that of \texttt{PACMAN} \cite{galluppi2012hierachical}, which is used to map SNNs on SpiNNaker neuromorphic hardware. Following are the limitations of \texttt{PACMAN} that motivated this paper. First, \texttt{PACMAN} is primarily targeted for architectures supported on SpiNNaker such as Deep Belief Networks \cite{stromatias2015scalable} and Convolution Neural Networks \cite{serrano2015convnets}. \texttt{PACMAN} offers limited flexibility to implement evolving computation models such as the Liquid State Machine (LSM) \cite{maass2002real} or Hierarchical Temporal Memory (HTM) \cite{hawkins2010hierarchical}. Second, \texttt{PACMAN} requires significant modification to support clustered neuromorphic architectures with local and global synapses. This is because the \texttt{PACMAN} tool is natively developed for SpiNNaker hardware only, where computing elements are ARM cores with conventional Von-Neumann architecture. Third, \texttt{PACMAN} determines neuron mapping without considering spike latency related performance distortions and interconnect energy consumption.

We propose a systematic approach to map trained SNNs on a neuromorphic hardware. Fundamental to this is our instantiation of Particle Swarm Optimization (PSO) \cite{eberhart1995new}, which partitions a given SNN into local and global synapses. Local synapses are mapped on fully connected crossbars and global synapses on time-multiplexed interconnect between the crossbars. The objective of this optimization is to minimize spike communication on the time-multiplexed interconnect, saving energy and improving performance (such as accuracy) of the overlaying application by reducing spike disorder count and inter-spike interval distortion. Our approach can be used with simulators as well as with real hardware. Although energy consumption for CxQuad hardware is used to demonstrate the power/performance improvement, our conceptual approach is however, generic and can be used for a range of devices with memristor-based synaptic elements.

\noindent \textbf{Contributions:} Following are our novel contributions
\begin{itemize}
	\item a systematic framework for partitioning and mapping a trained SNN on neuromorphic hardware;
	\item introduction of performance metrics for mapping SNNs on neuromorphic hardware;
	\item a PSO-based partitioning of SNN into local and global synapses to reduce spike communication and congestion on time-multiplexed interconnect;
	\item a Python-based open-source framework for simulating SNN on neuromorphic hardware; and
	\item thorough experimentation with realistic applications, demonstrating the advantage of our proposed approach.
\end{itemize}

We build our framework in Python interfacing \texttt{CARLsim}~\cite{beyeler2015carlsim}, a GPU-accelerated application-level spiking neural network simulator with an extended version of \texttt{Noxim}~\cite{catania2015noxim}, for simulating time-multiplexed interconnect. Results demonstrate the energy, latency and performance gains in the global communication network using our approach. The remainder of this paper is organized as follows. Neuromorphic platform description is provided in Section~\ref{sec:hardware} together with the introduced metrics. PSO-based partitioning is introduced next in Section~\ref{sec:pso}. Our proposed systematic framework is described in Section~\ref{sec:framework}. Results and discussions are provided in Section~\ref{sec:results}. Finally, the paper is concluded with future outlook in Section~\ref{sec:conclusions}.

\begin{figure}[t]
	\centering
	\centerline{\includegraphics[width=0.99\columnwidth]{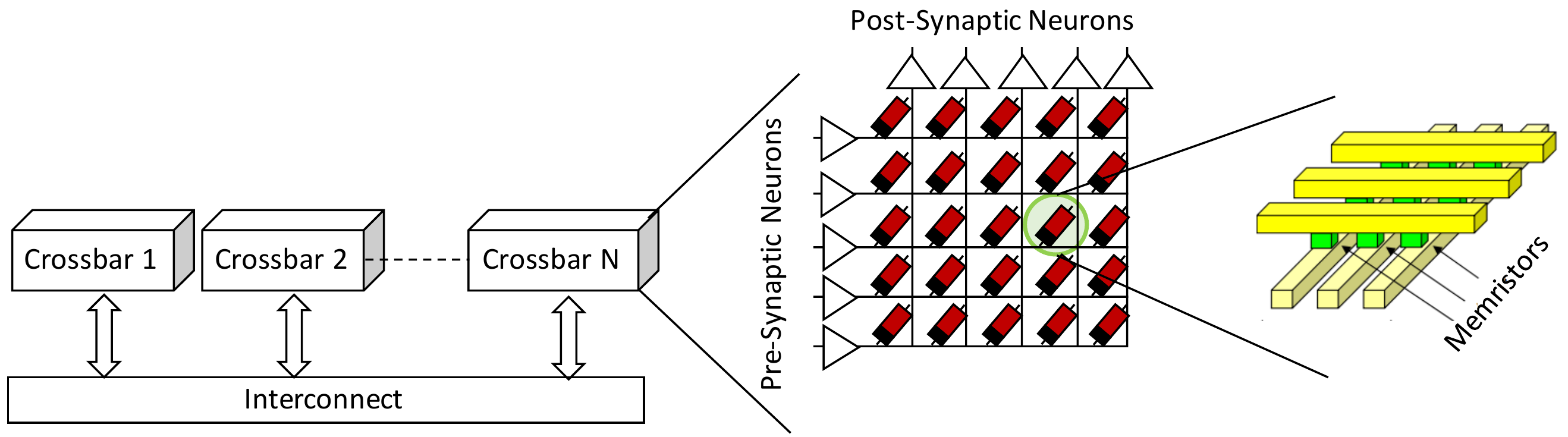}}
	\caption{A reference neuromorphic hardware}
	\label{fig:hw}
\end{figure}

\section{Neuromorphic Hardware and Related Metrics}
\label{sec:hardware}
Figure \ref{fig:hw} shows the general architecture of a modern neuromorphic hardware. The architecture consists of multiple crossbars of fully connected neurons (shown in part b). Implementation wise, a crossbar is a 3D arrangement of nanowires, with pre-synaptic neurons connected to bottom nanowires and post-synaptic neurons to the top nanowires (or vice-verse). Each crosspoint of a top and bottom nanowire, is a two-terminal memristor nanodevice (shown in part c). The synaptic connection strength between a pair of pre- and post-synaptic neurons is encoded as resistance value of the memristor connecting them and is adjusted by regulating the flow of current through it. Crossbars communicate with each other using an interconnect (shown in part a). Traffic on the interconnect is time-multiplexed. Several interconnect alternatives have been explored in literature for neuromorphic computing. The commonly used ones are NoC-tree (CxQuad) and NoC-mesh (TrueNorth, HiCANN).

Analogous to mammalian brain, synapses of a SNN can be classified into local and global synapses based on the distance information (spike) is conveyed. Local synapses are short distance links, where pre- and post-synaptic neurons are located in the vicinity. Global synapses are those where pre- and post-synaptic neurons are farther apart. To reduce power consumption of the hardware implementation of SNNs, two principles are widely adopted in the community:
\begin{itemize}
	\item the number of point-to-point local synapses is limited to a reasonable dimension (size of a crossbar); and
	\item instead of point-to-point global synapses (which are of long distance) as found in a mammalian brain, the hardware implementation usually consists of time-multiplexed interconnect shared between global synapses.
\end{itemize}

\begin{figure}[t]
	\centering
	\centerline{\includegraphics[width=0.9\columnwidth]{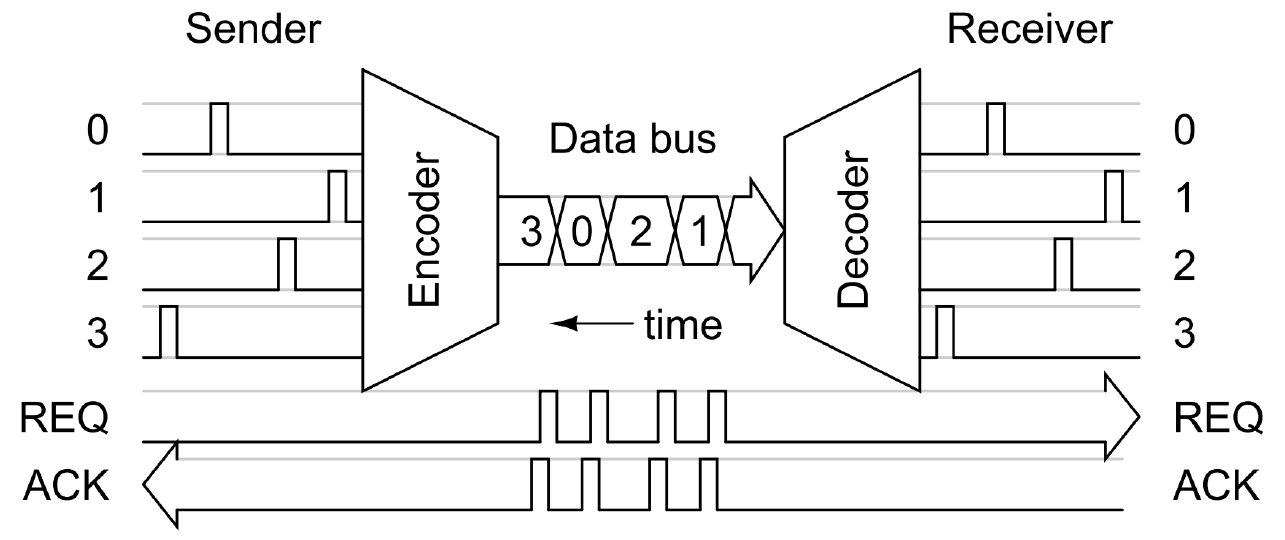}}
	\caption{An example showing AER protocol (adapted from \cite{boahen1998communicating}).}
	\label{fig:aer}
\end{figure}

CxQuad for example, consists of four crossbars, each with 128 pre- and 128 post-synaptic neurons implementing a full 16K (128x128) local synapses per crossbar. The crossbars are interconnected using a NoC-tree, which time-multiplexes global synaptic connections. The spike communication protocol for the global synapse interconnect is \texttt{Address Event Representation} (AER) \cite{boahen1998communicating}. An example is shown in Figure \ref{fig:aer} to explain the principles behind AER. Here four neurons in the input group (a crossbar) spikes at time 3, 0, 1 and 2 time units, respectively. The encoder encodes these four spikes in order to be communicated on the global synapse interconnect.  As can be clearly seen from this figure, a spike is encoded uniquely on the global synapse interconnect in terms of its source and time of spike.

\noindent Following are metric for global synapse interconnect.
\begin{itemize}
	\item Conventional metric for the global synapse interconnect
	\begin{itemize}
		\item \emph{Latency}: The delay between transmitting a spike packet (AER) by the encoder and receiving the packet by the decoder.
		\item \emph{Energy}: The energy consumed by spike communication on the global synapse interconnect.
	\end{itemize}
	\item Introduced metric for SNN performance on hardware
	\begin{itemize}
		\item \emph{Spike disorder count}: This is a measure of information loss in a SNN and is calculated based on the difference in spike order between sender and receiver neurons. An example is provided to explain this. Let us assume that neuron A and B need to communicate spikes to neuron C with spikes from A to be received before the spike from neuron B. Also let A, B and C are all mapped to different crossbars. If it happens that the crossbar with B is arbitrated to occupy the interconnect prior to crossbar with A, spikes from B will be received at C before the spike from A causing a disorder of spikes and potential information loss.
		\item \emph{Inter-spike distortion}: This is a measure for information distortion in temporally coded SNN and is measured by the difference between the sender neuron's inter-spike interval and receiver neuron's inter-spike intervals. Inter-spike distortion is attributed to spike congestion on the global synapse interconnect, causing some spike packets to be delayed than others.
	\end{itemize}
\end{itemize}

Based on the above discussions, the problem we aim to solve in this paper is as follows. Partition a given SNN into local and global synapses, where local synapses are mapped on the crossbar and global synapses on the global synapse interconnect. The objective of this optimization problem is to reduce spike communication (congestion) on the interconnect, which minimizes interconnect energy consumption, spike latency, spike disorder count and inter-spike distortion. 

\section{PSO-based Mapping of Trained SNN on Neuromorphic Hardware}
\label{sec:pso}
In this work we propose to use evolutionary techniques in order to solve the optimization problem of reducing spike communication on the interconnect, saving energy and improving SNN performance. We use particle swarm optimization (PSO) \cite{eberhart1995new}, an evolutionary computing technique inspired by social behaviors such as bird flocking and fish schooling. Evolutionary computing techniques are efficient in avoiding to stuck at local optima. Additionally, PSO is computationally less expensive with faster convergence compared to its counterparts such as genetic algorithm (GA) or simulated annealing (SA).

\begin{figure}[t]
	\centering
	\centerline{\includegraphics[width=0.99\columnwidth]{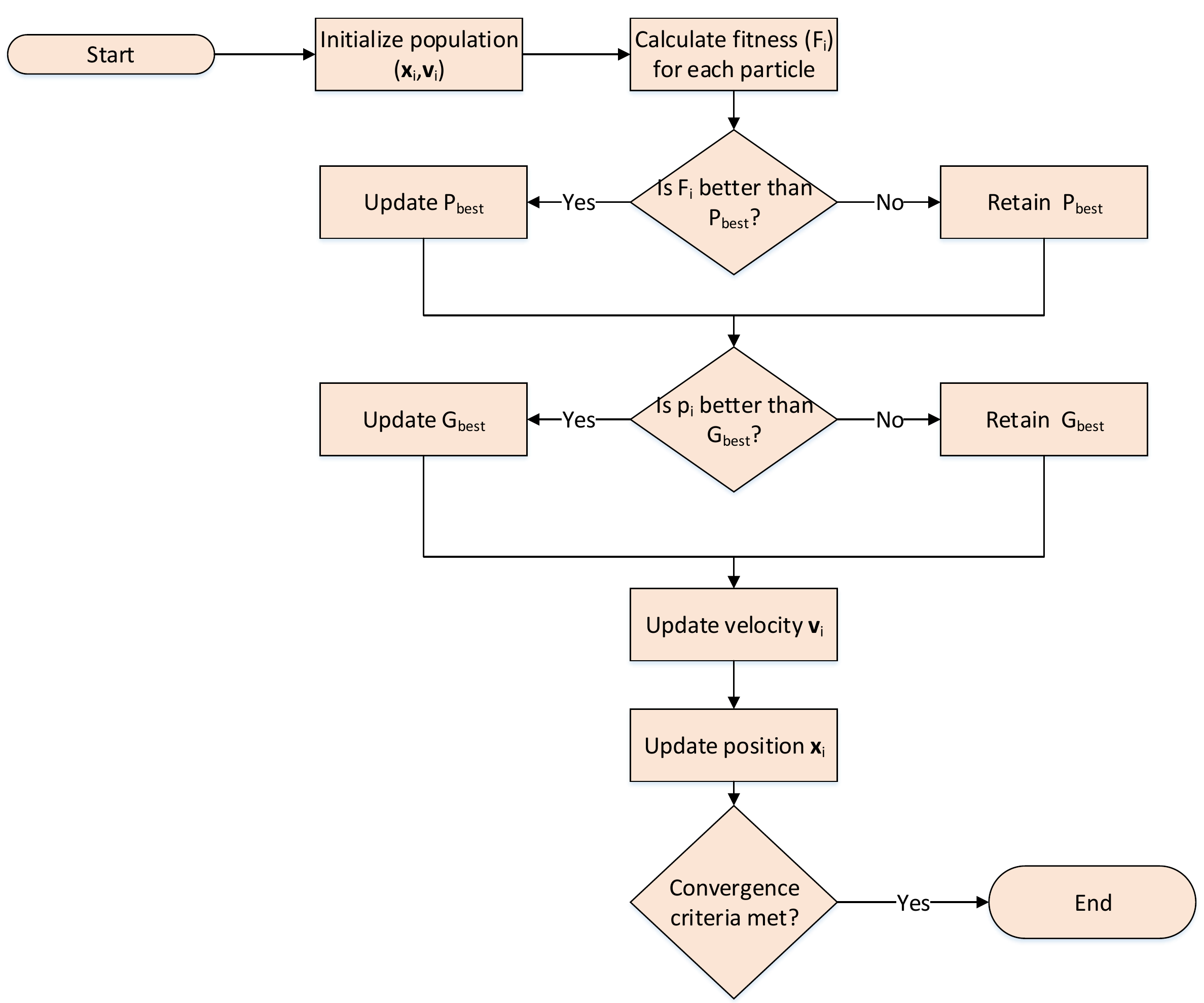}}
	\caption{Particle swarm optimization to find an optimum solution.}
	\label{fig:pso}
\end{figure}

In general, PSO finds the optimum solution to a fitness function $F$. Each solution is represented as a particle in the swarm. Each particle has a velocity with which it moves in the search space to find the optimum solution. During the movement, a particle updates its position and velocity according to its own experience (closeness to the optimum) and also experience of its neighbors. We introduce the following notations for PSO.

\begin{footnotesize}
\begin{align*}
D &= \text{dimensions of the search space}\\
n_p &= \text{number of particles in the swarm}\\
\mathbf{\Theta} = \{\mathbf{\theta}_l\in\mathbb{R}^{D}\}_{l=0}^{n_p-1} &= \text{positions of particles in the swarm}\\
\mathbf{V} = \{\mathbf{v}_l\in\mathbb{R}^{D}\}_{l=0}^{n_p-1} &= \text{velocity of particles in the swarm}
\end{align*}
\end{footnotesize}
\normalsize Position and velocity updates are performed according to

\begin{footnotesize}
\begin{align}
\label{eq:eq101}
\mathbf{\Theta}(t+1) &= \mathbf{\Theta}(t) + \mathbf{V}(t+1)\\
\mathbf{V}(t+1) &= \mathbf{V}(t) + \varphi_1\cdot\Big(P_{\text{best}}-\mathbf{\Theta}(t)\Big) + \varphi_2\cdot\Big(G_{\text{best}}-\mathbf{\Theta}(t)\Big)\nonumber
\end{align}
\end{footnotesize}
\normalsize where $t$ is the iteration number, $\varphi_1,\varphi_2$ are constants and $P_{\text{best}}$ (and $G_{\text{best}}$) is the particles own (and neighbors) experience. Figure \ref{fig:pso} shows the iterative solution of PSO, where position and velocity are updated until a predefined convergence is achieved. In the following we describe the transformation of our partitioning problem to the PSO domain.

Let us consider a SNN with $N$ neurons organized in a topology, specific to a given application (more details in Section \ref{sec:results}). The SNN can be represented as a graph $\mathcal{G} = (A,S)$, where $A = \{\mathbf{a_i}~|~0\leq i\leq N-1\}$ is the set of nodes of the graph, representing neurons and $S = \{s_{i,j}\}$ is the set of synapses representing connections between the neurons. Each synapse $\mathbf{s_{i,j}}$ is a tuple $\langle \mathbf{a_i},\mathbf{a_j},\mathbf{T_{i,j}}\rangle$, where $\mathbf{a_i}$ is the pre-synaptic neuron, $\mathbf{a_j}$ is the post-synaptic neuron and $\mathbf{T_{i,j}} = \{t_{i}^l~|~0\leq l \leq L_i \}$ are the spike times of the pre-synaptic neuron $\mathbf{a_i}$. This graph represents initial specification of a trained SNN in terms of synaptic weights and spike times. This graph is generated from \texttt{CARLsim} \cite{beyeler2015carlsim}.

Let $C$ be the number of crossbars in the architecture, with $N_c$ being the maximum number of neurons per crossbar. The specification $C$ is usually provided by a designer for a given architecture. In Section \ref{sec:results_dse} we discuss how to obtain this specification for a given set of applications. Let $x_{i,k} = \{0,1\}$, be the variable indicating if neuron $\mathbf{a_i}$ is allocated to crossbar $\mathbf{c_k}$, where $0\leq k\leq C$. These variables $x_{i,k}$'s are dimensions of our PSO with $D = N\cdot C$. In order to transform real-valued $x_{i,k}$ to binary values (for 0-1 assignments), the velocity and position updates need to be binarized. This is achieved using the following set of equations

\begin{footnotesize}
\begin{align}
\label{eq:binarization}
	&\hat{\mathbf{v}_{i,k}} = \texttt{sigmoid}(\mathbf{v}_{i,k}) = \frac{1}{1+\texttt{e}^{-\mathbf{v}_{i,k}}} = \begin{cases}
	0 \text{ if } \mathbf{v}_{i,k} < 0\\
	1 \text{ otherwise}
	\end{cases}\\
	&	\hat{x_{i,k}} = \begin{cases}
	0 \text{~~if } \texttt{rand()} < \hat{\mathbf{v}_{i,k}}\\
	1 \text{~~otherwise }
	\end{cases}
\end{align}
\end{footnotesize}
\normalsize Constraints for our PSO are as follows
\begin{itemize}
	\item every neuron is allocated to one crossbar only i.e.,
		\begin{equation}
		\label{eq:allocation_constraint}
		\footnotesize \sum_k \hat{x_{i,k}} = 1~~\forall i
		\end{equation}
	\item assignment must satisfy the dimensions of crossbar i.e.,
		\begin{equation}
		\label{eq:assignment_constraint}
		\footnotesize \sum_i \hat{x_{i,k}} \leq N_c~~\forall k
		\end{equation}
\end{itemize}

To determine the number of spikes communicated between crossbars $k_1$ and $k_2$, we define two sets $K_1$ and $K_2$, where $K_1$ ($K_2$) is the set of neurons allocated to crossbar $k_1$ ($k_2$). Elements of these sets are determined as follows
\begin{equation}
\label{eq:set}
\footnotesize K_1 = \{(i\cdot \hat{x_{i,k_1}})\}_{i=0}^{N-1} \text{ and } K_2 = \{(j\cdot \hat{x_{j,k_2}})\}_{j=0}^{N-1}
\end{equation}

The total spikes between crossbars $k_1$ and $k_2$ is
\begin{equation}
\footnotesize \text{spikes}(k_1,k_2) = \begin{cases}
0 & \text{ if } k_1 = k_2\\
\sum\limits_{\substack{i,j \\ i\in K_1 \\ j\in K_2}} \mathbf{T_{i,j}} & \text{ otherwise}
\end{cases}
\end{equation}


Total spikes on the global synapse interconnect is
\begin{equation}
\label{eq:gsi_spikes}
\footnotesize F = \sum_{k_1,k_2} \text{spikes}(k_1,k_2)
\end{equation}

The objective of the PSO is to minimize $F$ satisfying constraints Equation \ref{eq:allocation_constraint}--\ref{eq:assignment_constraint}. The assignment information (i.e., outcome of PSO) is used by the global synapse simulator (discussed in Section \ref{sec:framework}) to compute global statistics of spike communication. This statistics are then used to compute (1) the \texttt{spike disorder count} as the fraction of total spikes arriving out of order at the neurons and (2) the \texttt{Inter-spike distortion} as the maximum difference between the inter-spike interval of source and destination neurons.

\section{Systematic Partitioning Framework for SNNs}
\label{sec:framework}
\begin{figure}[t]
	\centering
	\centerline{\includegraphics[width=0.99\columnwidth]{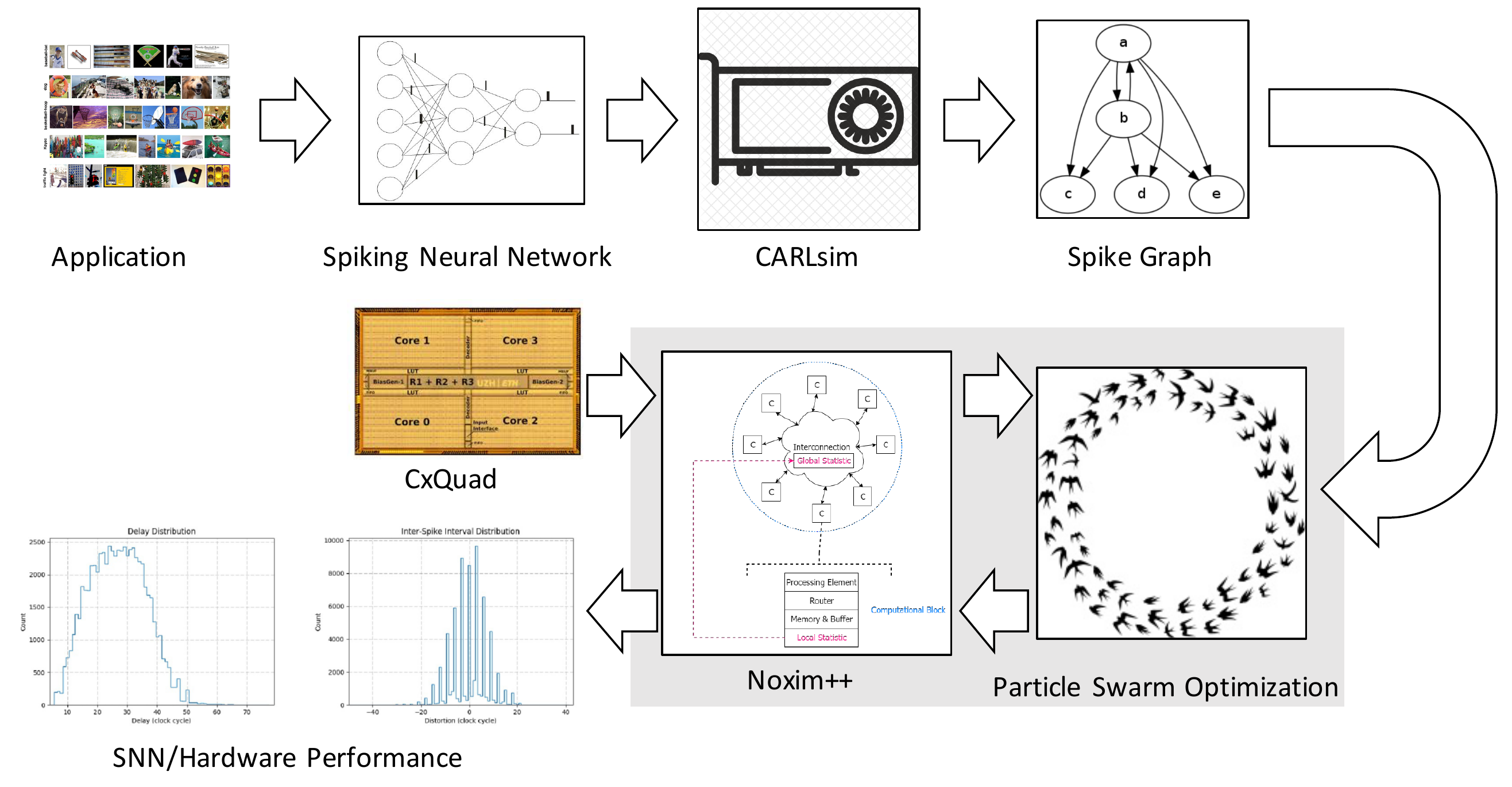}}
	\caption{Systematic partitioning of SNN on neuromorphic hardware.}
	\label{fig:framework}
\end{figure}

Figure \ref{fig:framework} shows our systematic framework. This framework takes an application implemented using SNN as input (shown at the top left). Table \ref{tab:nmc} provides a set of applications used to evaluate our partitioning methodology. It is to be noted that the first three applications are based on rate-coding, while the last one is with temporal coding. This provides a range of applications with conventional and evolving topologies with different spike coding schemes. 
Apart from these realistic applications, we also considered synthetic applications by varying the depth and width of SNN layers. 
All applications are first simulated using \texttt{CARLsim} \cite{beyeler2015carlsim}, a GPU-accelerated library for simulating spiking neural network models with a high degree of biological detail. 

The output of \texttt{CARLsim} is a trained SNN with spike times of each neuron in the SNN. This is converted into a dataflow graph as shown at the top right hand of Figure \ref{fig:framework}. The SNN graph (format is explained earlier) is presented to the PSO algorithm to partition into local and global synapses and map them on a neuromorphic hardware. The neuromorphic hardware architecture is input to \texttt{Noxim++}, which can incorporate details of a real chip (CxQuad in Figure \ref{fig:framework}). It is also possible to replace \texttt{Noxim++} with actual CxQuad or other neuromorphic chip directly, making the framework similar to \texttt{PACMAN}. The PSO formulation is discussed in the previous section.

The \texttt{Noxim++} simulator is an extended version of the originally proposed \texttt{Noxim} simulator \cite{catania2015noxim}, which is highly configurable network-on-chip simulation based on mesh architecture. The configurable parameters include buffer size, network size, packet size, packet injection rate, routing algorithm, selection strategy, among others. For the power consumption simulation, users can modify the power values in external loaded YAML file to benefit from the flexibility. During a simulation, \texttt{Noxim} calculates latency, throughput and power consumption automatically based on the statistics collected during runtime. The original \texttt{Noxim} simulator is extended with the following features for our framework
\begin{itemize}
	\item addition of different interconnect models for representative neuromorphic hardware
	\item incorporation of SNN-related metrics (spike disorder count and inter-spike distortion)
	\item multicast feature, where spike packets can be communicated to a selected subset of crossbars.
\end{itemize}

\begin{table}[t]
	\renewcommand{\arraystretch}{1.2}
	\caption{Realistic applications used for evaluating our approach.}
	\label{tab:nmc}
	\begin{scriptsize}
		\begin{center}
			\begin{tabular}{|l|l|l|}
				\hline
				Approach & Application & Topology\\
				\hline
				CARLsim native \cite{beyeler2015carlsim} & hello world (HW) & Feedforward (117, 9)\\
				CARLsim native \cite{beyeler2015carlsim} & image smoothing (IS) & Feedforward (1024, 1024)\\
				Diehl et al. \cite{diehl2015unsupervised} & handwritten digit (HD) & Unsupervised, recurrent (250, 250)\\
				Das et al. \cite{das2017unsupervised} & heartbeat estimation (HE) & Unsupervised, LSM (64, 16)\\
				\hline
			\end{tabular}
		\end{center}
	\end{scriptsize}
\end{table}

Overall, the framework generates different results, a snapshot of which is presented to the bottom left corner of Figure \ref{fig:framework}. The full framework will be released upon acceptance of this work for the benefit of research community.



\section{Results and Discussions}
\label{sec:results}
All experiments are conducted on Google Cloud Platform configured with 4 CPUs, 26 GB RAM and NVIDIA Tesla K80 GPU. The platform runs Ubuntu 14.04. Apart from the realistic applications of Table \ref{tab:nmc}, we considered synthetic applications with different number of neural network layers and number of neurons per layer. Our systematic framework is compared with \texttt{PACMAN} \cite{galluppi2012hierachical}, adapted for CxQuad architecture. Additionally, we also compare our approach with the ad-hoc mapping technique \texttt{NEUTRAMS} \cite{ji2016neutrams}, which uses a Network-on-Chip simulator to determine energy consumption on a neuromorphic architecture, without solving the local and global synapse partitioning problem and incorporating SNN performance.

\begin{figure}[t]
	\centering
	\centerline{\includegraphics[width=0.99\columnwidth]{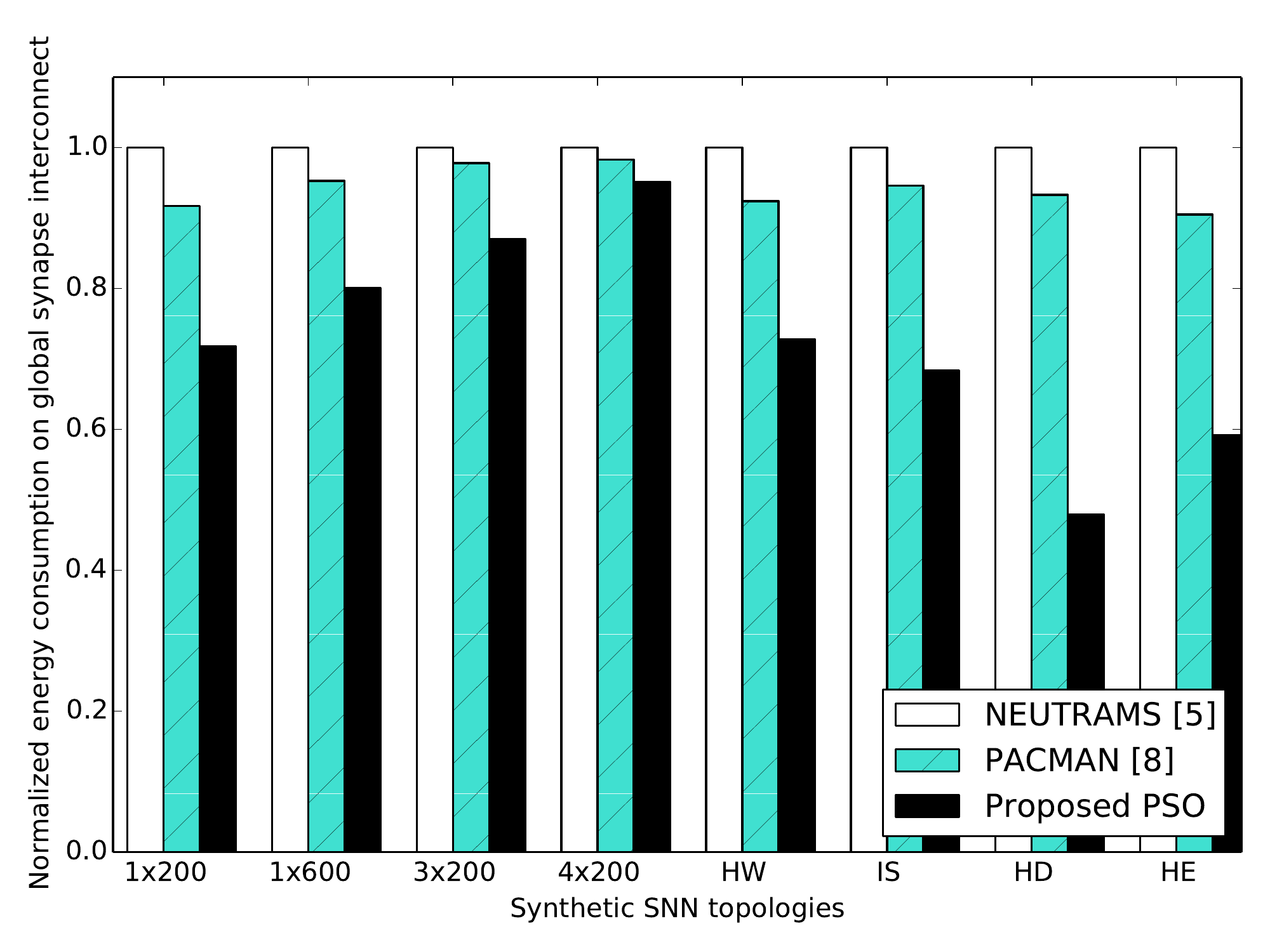}}
	\caption{Exploration with synthetic and realistic SNN-based applications.}
	\label{fig:synth}
\end{figure}



\subsection{Energy Comparison on Global Synapse Interconnect}
Figure \ref{fig:synth} reports the energy consumption on the global synapse interconnect for three approaches -- \texttt{NEUTRAMS} \cite{ji2016neutrams}, \texttt{PACMAN}~\cite{galluppi2012hierachical} and our proposed PSO-based partitioning. We evaluated 8 synthetic topologies (4 are plotted in Figure \ref{fig:synth}) and are marked on the X-axis with $m\times n$, where $m$ is the number of layers and $n$ is the number of neurons per layer. Neurons of the first layer in each of these topologies receive their input from 10 neurons creating spike trains, whose inter-spike interval follows a Poisson process with mean firing rates between 10 Hz and 100 Hz. Additionally, these synthetic SNNs implement fully connected feedforward topologies. Energy numbers for each topology are normalized with respect to \texttt{NEUTRAMS}. It can be seen clearly from Figure \ref{fig:synth}, that our proposed PSO-based partitioning achieves the minimum energy out of the three techniques. The improvement with respect to \texttt{NEUTRAMS} is between 2.4\% and 48.7\% (average 20.2\%), while that with respect to \texttt{PACMAN} is between 1.5\% and 45.4\% (average 17.2\%). It is to be noted that the energy improvement decreases with increase in the number of synapses. This is observed from results for topology 4x200 (with dense 122000 synapses) where energy consumption for the three approaches is comparable (energy gains are less than 2\%). For topology 1x200 (with 2000 synapses), improvements using our approach is more than 40\%. These results indicate that our approach is able to find the best partition for sparse and dense synapses, with higher improvements for sparse connections. Energy gains observed for the four realistic applications are in the range (27.0\% -- 52.1\%, average 38\%) with respect to \texttt{NEUTRAMS} and (21.2\% -- 48.7\%, average 33\%) with respect to \texttt{PACMAN}.

\begin{table}[t]
	\renewcommand{\arraystretch}{1.2}
	\caption{Metric evaluation for realistic applications.}
	\label{tab:snn_metric}
	\begin{scriptsize}
		\begin{center}
			\begin{tabular}{|l|c|c|c|c|}
				\hline
				\multirow{ 2}{*}{Metric} & \multicolumn{2}{|c|}{\texttt{hello\_world}}  & \multicolumn{2}{|c|}{\texttt{image smoothing}}\\
				\cline{2-5}
				& PACMAN \cite{galluppi2012hierachical} & Proposed & PACMAN \cite{galluppi2012hierachical} & Proposed\\
				\hline
				ISI Distortion (cycles) & 6.6 & 2.9 & 8.1 & 4.7\\
				Disorder count (\%) & 0.08 & 0.05 & 0.32 & 0.06\\
				Throughput (AER/ms) & 0.2 & 0.16 & 0.4 & 0.3\\
				Latency (cycles) & 108 & 70 & 99 & 69\\
				\hline
				\multirow{ 2}{*}{Metric} & \multicolumn{2}{|c|}{\texttt{digit recog.}}  & \multicolumn{2}{|c|}{\texttt{heartbeat estimation}}\\
				\cline{2-5}
				& PACMAN \cite{galluppi2012hierachical} & Proposed & PACMAN \cite{galluppi2012hierachical} & Proposed\\
				\hline
				ISI Distortion (cycles) & 6.2 & 4.3 & 4.9 & 3.9\\
				Disorder count (\%) & 0.02 & 0.01 & 0.12 & 0.02\\
				Throughput (AER/ms) & 0.5 & 0.4 & 0.3 & 0.33\\
				Latency (cycles) & 150 & 146 & 216 & 171\\
				\hline
			\end{tabular}
		\end{center}
	\end{scriptsize}
\end{table}

\subsection{SNN Metric Evaluation on Global Synapse Interconnect}
Table \ref{tab:snn_metric} reports results for other metrics relevant to SNNs for realistic applications. Specifically, the table reports average inter-spike interval (ISI) distortion in terms of interconnect clock cycles in rows 3 \& 9; the spike disorder count as a fraction of the total spikes in rows 4 \& 10; the average throughput in terms of number of AER packets per ms on the global synapse interconnect in rows 5 \& 11; and finally the maximum latency of spike communication on the global synapse interconnect in terms of interconnect clock cycles in rows 6 \& 12. Furthermore, the table compare results from our approach with that from \texttt{PACMAN} \cite{galluppi2012hierachical}. As can be seen clearly from this table, our approach outperforms \texttt{PACMAN} in terms of ISI distortion achieving an average 37\% fewer global synapse interconnect cycles. It is to be noted that ISI distortion significantly impacts application performance with temporal information coding such as the \texttt{heartbeat estimation}, where we observe that 20\% reduction of ISI distortion improves estimation accuracy by over 5\%. For the other 3 applications which are based on rate coding, the accuracy improvement due to ISI distortion is insignificant. In terms of spike disorder count, we observe that our approach achieves an average 63\% lower spike arrival disorder compared to \texttt{PACMAN}. It is to be observed that the throughput in \texttt{PACMAN} is usually higher than that of our approach. This is because the number of spikes communicated on the global synapse interconnect is usually higher in \texttt{PACMAN}. Finally, the spike propagation latency on the global synapse interconnect is also lower using our approach by 22\% (2\% -- 35\%). Improvements are consistent for the 8 evaluated synthetic topologies. These improvements are due to our PSO-based partitioning, which reduces spike congestion on the global synapse interconnect, improving communication energy and latency.

\begin{figure}[t]
	\centering
	\centerline{\includegraphics[width=0.99\columnwidth]{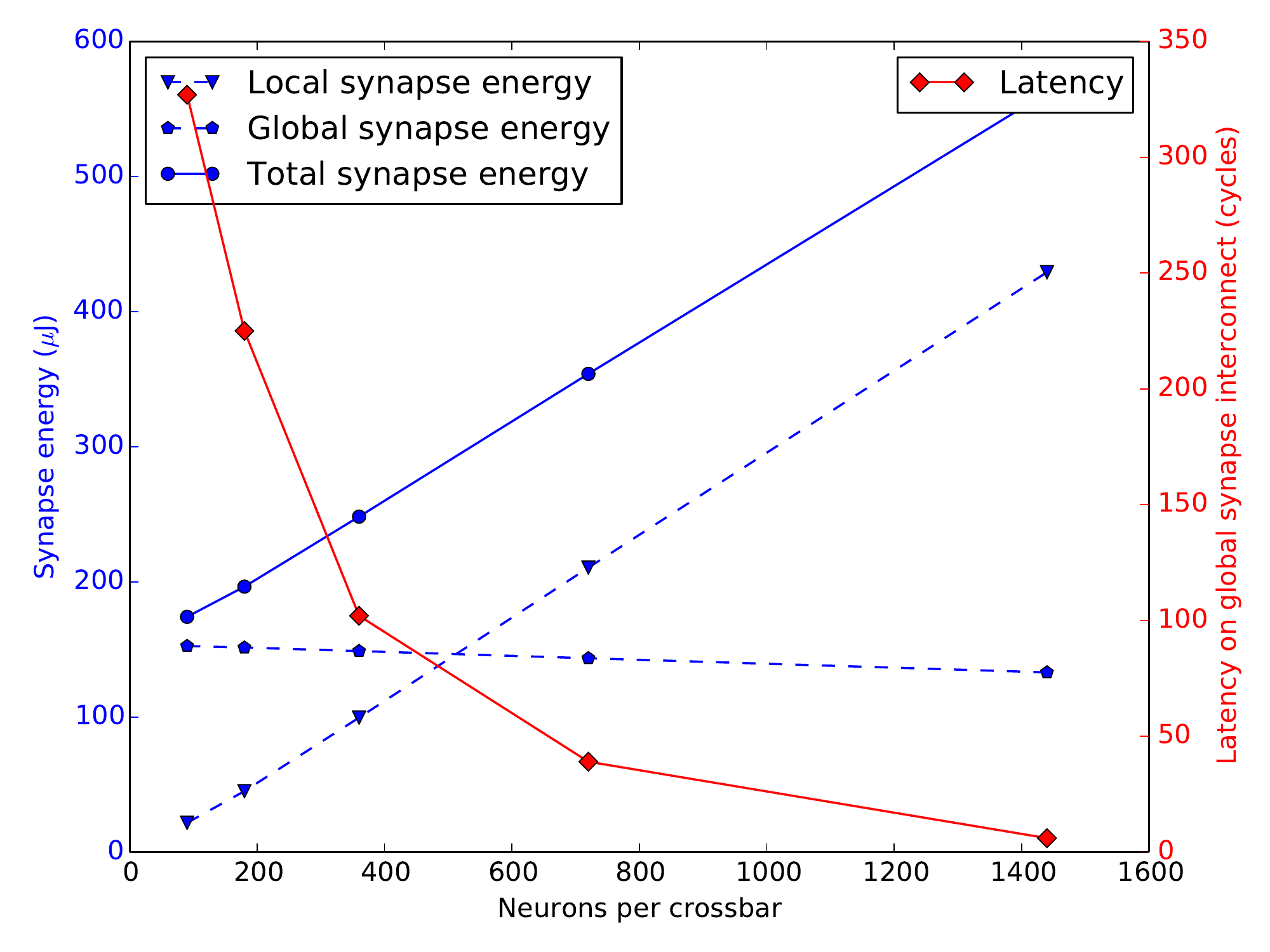}}
	\caption{Architecture exploration with hand-written digit recognition \cite{diehl2015unsupervised}.}
	\label{fig:arch}
\end{figure}

\subsection{Neuromorphic Architecture Exploration}
\label{sec:results_dse}
To further demonstrate the broad usage of our approach and framework, we take an application (\texttt{digit recognition} \cite{diehl2015unsupervised}) and explore architectural alternatives. We aim to answer the following question: given an application, which is preferred between an architecture with fewer number of large crossbars or an architecture with large number of small crossbars? Figure \ref{fig:arch} plots exploration results in terms of local/global synapse energies and latency for spike communication on the global synapse interconnect. The number of neurons per crossbar is increased from 90 to 1440. The local synapse energy is the total energy for spike communication inside all crossbars in the architecture. The global synapse energy is the total energy for spike communication on the global synapse interconnect. The energy numbers are averaged for processing/interpreting a hand-written digit image of 28x28 pixels. Latencies in this figure are worst-case values for all spikes communicated on the global synapse interconnect. 

Following observations can be made from Figure \ref{fig:arch}. First, as the size of crossbars increases, more synapses are mapped by our approach inside a crossbar reducing the number of spikes and hence energy on the global synapse interconnect. Conversely, the local synapse energy increases due to more spike communication inside a crossbar. Second, the worst-case spike latency on the global synapse interconnect decreases with increase in the size of crossbars. This is because, as more synapses are mapped locally, congestion on the interconnect reduces, reducing the latency. Clearly, the best choice is an intermediate point in between the extremes. So we clearly need our framework to explore this complex search space. Once the best solution is identified, the size (and hence the number) of crossbars for a given application can be determined, together with partitioning the underlying SNN into local and global synapses. The mapping can then be enforced on the CxQuad board at design-/configuration time.


\begin{figure}[t]
	\centering
	\centerline{\includegraphics[width=0.99\columnwidth]{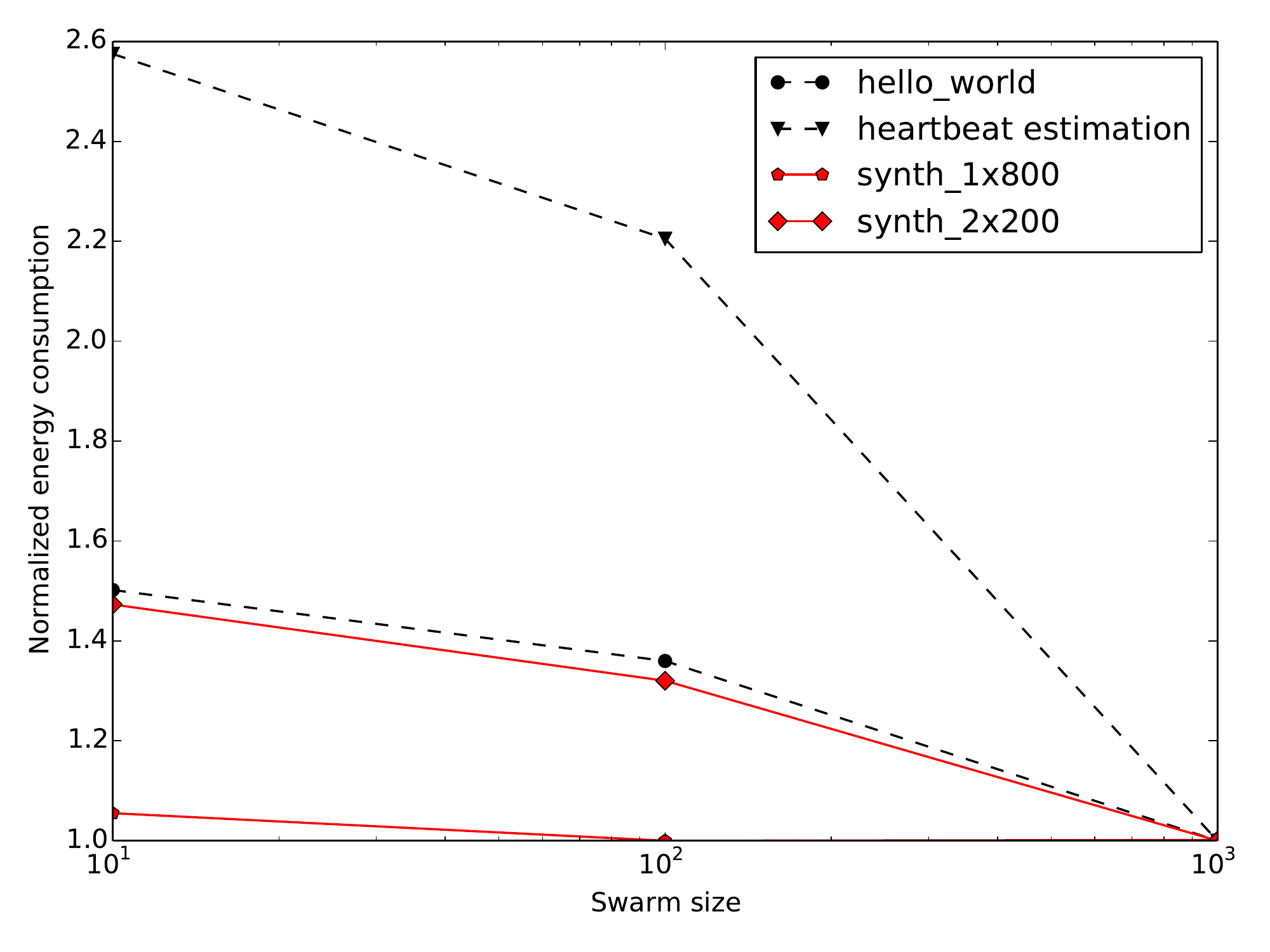}}
	\caption{Exploration with swarm size.}
	\label{fig:swarm_size}
\end{figure}

\subsection{PSO-Related Exploration Results}
In this section we present results related to our particle swarm optimization. Figure \ref{fig:swarm_size} plots the energy consumption on the global synapse interconnect with different number of swarm particles (in log scale) for four applications -- two real and two synthetic, with the number of iterations fixed to 100. Energy results are normalized with respect to the minimum energy obtained for these applications. Furthermore, there are no improvements in the energy consumption with more than 1000 particles. Hence the x-axis is limited to 1000 particles. As can be seen, with more number of particles, the algorithm is able to find better results for a fixed number of iteration of the algorithm. For applications \texttt{synth\_2x200}, the minimum is reached for a swarm size of 105. For other three applications, the point of minimum energy is close to 1000.  Trends are consistent for other applications considered. Based on these results, we used a swarm size of 1000 for our experiments. The average wall clock time to find an optimum solution using these settings (swarm size = 1000, iteration = 100) is average 35 minutes on the Google Cloud platform. Detailed timing results are omitted for space limitation.


\section{Conclusions and Outlook}
\label{sec:conclusions}
This paper presents an approach to map local and global synapses of a SNN-based application on a crossbar-based neuromorphic architecture. Fundamental to our approach is the use of particle swarm optimization, which partitions SNN-based applications, with local synapses mapped to crossbars and global synapses mapped on a time-multiplexed interconnect. Furthermore, we proposed two metrics -- inter-spike interval distortion and disorder count, specific to SNN performance for spatial and temporal information coding. Using realistic and synthetic applications we show that our approach reduces spike communication on the global synapse interconnect, reducing communication energy by an average 33\% and spike propagation latency by an average 22\%, with respect to \texttt{PACMAN} \cite{galluppi2012hierachical}, which is the standard SNN mapping technique for SpiNNaker. Run-time SNN mapping will be addressed in future.

\bibliographystyle{IEEEtran}
\bibliography{snnhw}


\end{document}